# A Decision Support System for Inbound Marketers: An Empirical Use of Latent Dirichlet Allocation Topic Model to Guide Infographic Designers

*Meisam Hejazi Nia*, University of Texas at Dallas

**ABSTRACT**
Infographic is a type of information presentation that inbound marketers use. I suggest a method that can allow the infographic designers to benchmark their design against the previous viral infographics to measure whether a given design decision can help or hurt the probability of the design becoming viral.

## Introduction

In contrast to outbound marketers that use direct paper mail, radio, TV advertising, and sales flyers, inbound marketers promote the company's brand through blogs, podcasts, video, e-books, e-newsletters, white papers, search engine optimization, and social media marketing. In a nutshell, inbound marketers concern about attracting the customers through creating a viral quality content. Inbound marketers that publish infographics grow their traffic an average of 12% more than those that do not.[1] The source of this improvement is in what Edward Tufte[2] suggests as high processing power of human vision, in contrast to the limited processing capacity of human mind, which can only analyze 5 to 9 processes simultaneously.

There is barely appropriate guide to suggest what design choices make an infographic viral. Given these benefits of the infographics and the fact that they have not been studied quantitatively yet, I ask the following questions: Can the low level feature of an infographic guide an infographic designer to design a viral infographic? Can I design a decision support system to allow an infographic designer to measure the effect of her design decisions on the probability of the infographic becoming viral? What are the current viral topics for which the practitioners create infographics?

To answer these questions, I develop a four staged machine learning pipeline, based on experimenting different approaches. I use an Optical Character Recognition (OCR) with a dictionary filter and the word-Net and the Google's word2vec to extract the verbal information of the infographics and k-mean to extract a histogram of five clusters of RGB and HSV of the images, to create a bag of verbal and visual words. Then I use a soft-clustering generative latent Dirichlet model (LDA), to identify twelve clusters of infographics that I labeled based on the word cloud of their titles. I applied my system on a data set of 355 infographics that I collected from Pinterest, Hubspot and information is beautiful websites. Based on the model free evidences, I find the infographics about world's top issues and the world's demographic has significantly higher social media hit than the social media and mobile infographics.

The method, I suggest, can allow the infographic designer to benchmark her design against the previous viral infographics to measure whether a given design decision can help or hurt the probability of the design becoming viral. The merit of my pipeline is its ability in summarizing big data (i.e.

---

[1] http://blog.hubspot.com/blog/tabid/6307/bid/33423/19-Reasons-You-Should-Include-Visual-Content-in-Your-Marketing-Data.aspx
[2] https://www.youtube.com/watch?v=g9Y4SxgfGCg

image of millions of pixels) into the predictive probability of an infographic being viral. The underlying assumption to my approach is that although infographic as an art work is unique, yet there are some common features of infographics in a form of underlying patterns that makes an infographic pleasant and viral. This assumption may be backed up by practitioners' suggestions to piggy back on the successful art-works to help the new art-work become viral.[3] Next, I will describe, the data, the method, the results, and the managerial implications with more detail.

## Data

My data includes 355 infographics with their social media activity that I collected from Pinterest, Hubspot, and information is beautiful website. The social media activity includes the total number of shares on social media websites, i.e. Facebook, Pinterest, Linkedin, and Twitter. Each infographic image includes millions of pixels with a triple of red, blue and green (RGB) color. I augmented this data with mapping of hue, saturation, value (HSV), as this scale is more perceptually relevant to the infographic audiences.[4] To summarize this matrix of million pixels I use an approach similar to Csurka et al. (2004) and Yang et al. (2007) to create a bag of visual words. In a nutshell, I cluster the pixels of the image into five clusters, and I sorted the density of pixels in each cluster, to create a bag of visual word. This approach is used in image compression, as an alternative to furrier transformation, to compress the image by keeping only relevant enough information in the image. An alternative approach is to use scale-invariant feature transform (or SIFT); however, although that approach may be relevant for object recognition, it is more complicated, and it may not keep enough information about the color of the image which is an important design element.

I also use an OCR engine augmented with an English word dictionary to extract the verbal information in each infographic. As each infographic is a sparse bag of verbal word, I use word-Net and Google's word2vec to capture similarity between the bags of verbal words of different infographics. WordNet and Google's word2vec are a lexical database for the English language. They group English words into sets of synonyms called synsets, provide short definitions and usage examples, and record a number of relations among these synonym sets or their members.[5] In the next step, I combined the full set of image and text features of info graphics into a single doc-term matrix. This matrix has 355 rows, i.e. for each document a separate row, and 392 columns. However, many of the elements of this matrix are zero, so the matrix is sparse. This scarcity suggested that I use dimensionality reduction techniques. As a result, I used Single Value Decomposition (SVD) method to keep 95% of the variation, and the output includes 30 new features for each infographic. Next I discuss related researches.

## Related Works

This work is related to studies by Siricharoen (2013), Milovanovic and Ivanisevic (2014), and Ma et al (2004) on infographics. Siricharoen (2013) suggests that infographics can be a marketing tool for e-entrepreneurship growth, for their well-organized easy-to-understand, eye-catchy, and sharability. Milovanovic and Ivanisevic (2014) study psychological factors including visual pictorial primacy, attracting and retaining attention, content organization, salability and astatic that make infographic as an effective widespread mean for communication.

This work is also related to the work by Blei et al (2003), Bishop (2006), and Hornik and Grün, (2011), from methodological point of view. Although these works have suggested the use of LDA method and incorporation of kernels at abstract level, my work tries to apply and integrate these approaches for the specific application of infographic design, and infographic designer decision support system. To the best of my knowledge this study is the first attempt to quantify the underlying features of viral infographics to build a decision support for infographic designers. From methodology standpoint, my approach gives a measure of the probability of membership in a cluster of viral infographics on each change that a designer makes. Next I will discuss the method I used.

In summary, practitioners list ten steps to create an effective infographic: (1) gathering data, (2) reading and highlighting facts, (3) finding the narrative, (4) identifying problems, (5) creating a hierarchy, (6) building a wireframe, (7) choosing a format, (8) determining a visual approach, (9) refinement and testing, and (10) releasing it into the world.[6] Design is an iterative process, and it requires refinement and testing. My approach uses machine learning approaches to extract collective wisdom of low level features (patterns) that create a viral infographic to guide designer in stage 8 and 9. In other words, I attempt to quantify the art of user acceptance in infographic design, by extracting low level features of viral infographics.

## Methodology and Estimation

I start this section with defining my four staged machine learning pipeline, or as I call them the stages of the process of infographic-design decision support system. Figure 1 shows my machine learning pipeline. I coded the process stages except the LDA stage in Python, and I used interfaces of WEKA for SVD and k-mean algorithms, and NLTK interface for wordNet and word2Vec. For Latent Dirichlet Allocation (LDA) I used R-interface of topic-models package.

---

[3] http://blog.hubspot.com/blog/tabid/6307/bid/33611/7-Companies-That-Jumped-on-a-Viral-Craze-at-Just-the-Right-Time.aspx
[4] http://en.wikipedia.org/wiki/HSL_and_HSV
[5] http://en.wikipedia.org/wiki/WordNet
[6] http://www.fastcodesign.com/1670019/10-steps-to-designing-an-amazing-infographic

I wanted to use a soft-clustering algorithm for histogram extraction from each image; however, Gaussian mixture model takes a long time to converge, so I used k-mean approach. To model the clusters of each infographic I used LDA. LDA is a generative structural approach to model the clusters of collection of items. This approach allows extracting membership probability of unseen examples based on the calibrated model. An LDA model, as a form of three layer hierarchical model, takes advantage of the membership information of the bag of words in each of the documents. In interest of saving space I refer interested readers to check Blei et al (2003) and Hornik and Grün (2011) for detail of LDA approach, and its advantages.

To estimate LDA model I define the likelihood of the model as follows:

$$p(D|\alpha,\beta) = \prod_{d=1}^{M} \int p(\theta_d|\alpha)(\prod_{n=1}^{N_d} \sum_{z_{d_n}} p(z_{d_n}|\theta_d) p(w_{d_n}|z_{d_n},\beta)) d\theta_d$$

The key inferential problem to solve for LDA is computing posterior distribution of topic hidden variables $\theta_d$, $z_d$, the first one with Dirichlet distribution, and the second one with multinomial distribution. To normalize the distribution of words given $\alpha$ and $\beta$ I marginalize over the hidden variables as follows:

$$p(D|\alpha,\beta) = \prod_{d=1}^{M} \frac{\Gamma(\sum_i \alpha_i)}{\prod_i \Gamma(\alpha_i)} \int (\prod_{i=1}^{k} \theta_i^{\alpha_i-1})(\prod_{n=1}^{N_d} \sum_{i=1}^{k} \prod_{j=1}^{V} (\theta_i \beta_{ij})^{w_n^j}) d\theta$$

Due to the coupling between $\theta$ and $\beta$ in the summation over latent topics this likelihood function is intractable. Therefore to estimate it Blei et al. (2003) suggests using variational inference method. Variational inference or variational Bayesian refers to a family of techniques for approximating intractable integrals arising in Bayesian inference and machine learning. An alternative approach is to use Gibbs sampling (Hornik and Grün, 2011). In appendix A, I present the Gibbs estimation procedure for LDA model. On a final note, I benchmarked CTM versus the LDA model, and the LDA method fit my data in terms of log likelihood better than CTM.

## Results

To evaluate the models I use log-likelihood to find the appropriate number of clusters, and to find appropriate features. In interest of space, I do not present the log likelihood of each model for different number of clusters, in this paper. However, the LDA approach that uses only the image data, i.e. RGB and HSV density of each cluster and the mean for the each infographic, and not the verbal information of the infographics fits my data better in terms of log likelihood. I initialized CTM model multiple times and selected the maximum of likelihood across iterations. This is because CTM uses Variational Expectation Maximization (VEM) algorithm, which has Expectation Maximization (EM) algorithm in its core, and EM algorithm is prone to the problem of multiple modes.

**Figure 1.** Machine Learning Pipeline for Infographic-Design Decision Support System

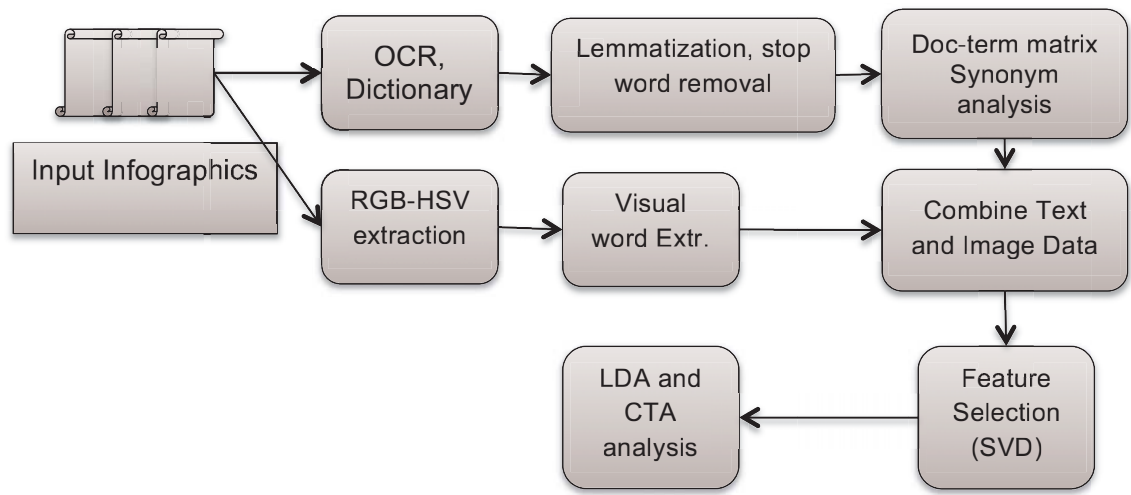

A model with twelve topics fit my data better, so I kept the assignment of twelve clusters. Table 1 presents the basic statistics of the infographic clusters that I identified with their labels. I used a word cloud for each cluster titles to name them. The word cloud arranges the keyword in a given corpus (i.e. here the collection of titles of infographics within each segment), so that the words that have higher frequency become bolder. The basic idea behind this approach is that each infographic per definition is supposed to be created around a central main point. In addition, infographic creators select the title for their infographic meticulously to make sure that both it reflects its content, and it is general enough to be picked up as a relevant link by the search engines.

Given the infographic clusters, I run a model free t-test to compare whether social media activities (i.e. number of shares on Facebook, Pinterest, Linkedin, Twitter) as a measure of infographic virality differ systematically across the infographics clusters. Table 2 presents the result of this between group t-tests. Based on this analysis I find that cool infographics about world's top issues and demographics has significantly higher social media hit than mobile and social media marketing infographics. In addition, infographics that contrast traditional and modern marketing approaches have significantly higher social media hits than the other infographics. Moreover, interactive marketing infographics have significantly higher social media hits than the social media marketing type infographics. Given these results I next discuss possible managerial implications of these results and the methods I employed.

## Managerial Implications

The result of this study may suggest that low level features of an infographic can systematically cluster different infographics into viral and non-viral infographics, so the proposed model may help the infographic designers to measure the impact of each design decision on the probability of their infographic becoming viral. To fulfill such task, my decision support system (DSS) first extracts the vector of visual words of an infographic through a k-mean algorithm. Then it uses the calibrated LDA engine to find the probability of the membership of the given infographic into each of the clusters. Given these probabilities, my system calculates the expected level of social media activities for a given infographic. In addition, my DSS gives the infographic designer the probability that the designed infographic can become viral, through presenting the probability that the designed infographic is the member of the viral infographic clusters. As a result, an infographic designer may be able to use the design principles to create an infographic, and she can measure the effect of each of her design decisions on the probability that her infographic becomes viral.

## Managerial Implications and Conclusion

In this study, I use a set of 355 infographics that I have collected from various websites including: Pinterest, hubspot, and informationisbeautiful.net, to quantify features that make an effective infographic. To do so, I use a four staged machine learning approach. To extract image information, in the first step I use RGB and HSV information of pixels of an infographic to create a vector of visual words. To extract the vector of visual words, I use an k-mean algorithm to identify five clusters in each image, and I build sorted histogram of the RGB and HSV information of each image. To extract text information, I also use an OCR combined with a dictionary process to extract text within the infographics. I merge both verbal and visual word vectors next and run two soft clustering methods, i.e. Latent Dirichlet Allocation

**Table 1.** The Basic Statistics of Clusters Social Media Activity

| Cluster ID | Cluster Name | Frequency | Average | Variance |
| --- | --- | --- | --- | --- |
| 1 | Cool info graphics about world's demographic infographics | 28 | 2303.143 | 8744003 |
| 2 | Mobile and Buzz Design Infographics | 30 | 923.5333 | 1904941 |
| 3 | Marketing design and Dashboard Infographics | 53 | 1254.528 | 3987451 |
| 4 | Face and Media Infographics | 9 | 446.6667 | 350812.4 |
| 5 | Traditional Marketing Infographics | 31 | 2693.032 | 10011501 |
| 6 | Social Media and Decision Making Infographics | 26 | 960.1538 | 869841.9 |
| 7 | General life Infographics | 39 | 1774.615 | 5735747 |
| 8 | Online professional design Infographics | 33 | 1414.455 | 5010189 |
| 9 | Responsive logos and brands Infographics | 15 | 1194.6 | 3101275 |
| 10 | International and online design Infographics | 35 | 1354.057 | 6700740 |
| 11 | Interactive Marketing Infographics | 28 | 1030.571 | 5468611 |
| 12 | Traditional vs. Online Media Infographics | 28 | 1717.643 | 7377299 |

Table 2. Comparing Viral Measure (Social Media Activity) of Clusters Together by Pairwise t-Test (the First Element Represents the Between Group t-Stat and the Second Element Is Corresponding t-Test Critical Value)

| | Cluster 2 | Cluster 3 | Cluster 4 | Cluster 5 | Cluster 6 | Cluster 7 | Cluster 8 | Cluster 9 | Cluster 10 | Cluster 11 | Cluster 12 |
|---|---|---|---|---|---|---|---|---|---|---|---|
| Cluster 1 | (2.3, 2)* | (1.89, 1.99)* | (1.85, 2.03)* | (−0.49, 2) | (2.21, 2)* | (0.81, 2) | (1.33, 2) | (1.33, 2.02) | (1.36, 2) | (1.79, 2) | (0.77, 2) |
| Cluster 2 | | (−0.8, 1.99) | (1, 2.02) | (−2.81, 2)* | (−0.11, 2) | (−1.74, 1.99) | (−1.04, 2) | (−0.57, 2.01) | (−0.82, 2) | (−0.21, 2) | (−1.42, 2) |
| Cluster 3 | | | (1.2, 2) | (−2.56, 1.99)* | (0.71, 1.99) | (−1.13, 1.99) | (−0.34, 1.99) | (0.11, 2) | (−0.2, 1.99) | (0.45, 1.99) | (−0.87, 1.99) |
| Cluster 4 | | | | (−2.1, 2.02)* | (−1.54, 2.03) | (−1.64, 2.01) | (−1.27, 2.02) | (−1.22, 2.06) | (−1.04, 2.02) | (−0.73, 2.03) | (−1.38, 2.03) |
| Cluster 5 | | | | | (2.69, 2)* | (1.38, 1.99) | (1.88, 2) | (1.7, 2.01) | (1.89, 2) | (2.27, 2) | (1.26, 2) |
| Cluster 6 | | | | | | (−1.65, 2) | (−0.97, 2) | (−0.56, 2.02) | (−0.74, 2) | (−0.14, 2) | (−1.35, 2) |
| Cluster 7 | | | | | | | (0.66, 1.99) | (0.85, 2) | (0.73, 1.99) | (1.27, 2) | (0.09, 2) |
| Cluster 8 | | | | | | | | (0.34, 2.01) | (0.1, 2) | (0.65, 2) | (−0.48, 2) |
| Cluster 9 | | | | | | | | | (−0.22, 2.01) | (0.24, 2.02) | (−0.67, 2.02) |
| Cluster 10 | | | | | | | | | | (0.51, 2) | (−0.54, 2) |
| Cluster 11 | | | | | | | | | | | (−1.01, 2) |

*Indicates whether the difference is significant with for 95% confidence interval.

(LDA), and Correlated Topic Model (CTM) to cluster the infographics. I identified twelve different clusters of infographics. I named the clusters based on the word cloud of labels of infographics items within the clusters. Also based on model free evidences I find that cool infographics about world's top issues and demographics has significantly higher social media hit than mobile and social media marketing infographics. In addition, infographics that contrast traditional and modern marketing approaches have significantly higher social media hits than other demographics. From methodology standpoint, my approach gives a measure of the probability of membership in a cluster of viral infographics on each change that a designer makes. Next step involves showing predictive validity of the proposed approach.